\documentstyle[11pt,amssymb,epsf,cite]{article}

\makeatletter
\@addtoreset{equation}{section}
\makeatother

\def\ben{\begin{equation}}
\def\een{\end{equation}}

\let\a=\alpha    
    
  \let\n=\nu

\let\X=\Xi     
\let\C=\Chi

\def\nn{\nonumber} \def\bd{\begin{document}} \def\ed{\end{document}}
\def\ds{\documentstyle} \let\fr=\frac \let\bl=\bigl \let\br=\bigr
\let\Br=\Bigr \let\Bl=\Bigl
\let\bm=\bibitem
\let\na=\nabla
\let\pa=\partial \let\ov=\overline
\newcommand{\be}{\begin{equation}}
\newcommand{\ee}{\end{equation}}
\def\ba{\begin{array}}
\def\ea{\end{array}}
\def\ft#1#2{{\textstyle{{\scriptstyle #1}\over {\scriptstyle #2}}}}
\def\fft#1#2{{#1 \over #2}}
\def\del{\partial}
\def\vp{\varphi}
\def\sst#1{{\scriptscriptstyle #1}}
\def\oneone{\rlap 1\mkern4mu{\rm l}}
\def\td{\tilde}
\def\wtd{\widetilde}
\def\ie{\rm i.e.\ }
\def\dalemb#1#2{{\vbox{\hrule height .#2pt
        \hbox{\vrule width.#2pt height#1pt \kern#1pt
                \vrule width.#2pt}
        \hrule height.#2pt}}}
\def\square{\mathord{\dalemb{6.8}{7}\hbox{\hskip1pt}}}
\newcommand{\ho}[1]{$\, ^{#1}$}
\newcommand{\hoch}[1]{$\, ^{#1}$}
\newcommand{\bea}{\begin{eqnarray}}
\newcommand{\eea}{\end{eqnarray}}
\newcommand{\ra}{\rightarrow}
\newcommand{\lra}{\longrightarrow}
\newcommand{\Lra}{\Leftrightarrow}
\newcommand{\ap}{\alpha^\prime}
\newcommand{\bp}{\tilde \beta^\prime}
\newcommand{\tr}{{\rm tr} }
\newcommand{\Tr}{{\rm Tr} }
\def\0{{\sst{(0)}}}
\def\1{{\sst{(1)}}}
\def\2{{\sst{(2)}}}
\def\3{{\sst{(3)}}}
\def\4{{\sst{(4)}}}
\def\5{{\sst{(5)}}}
\def\6{{\sst{(6)}}}
\def\7{{\sst{(7)}}}
\def\8{{\sst{(8)}}}
\def\n{{\sst{(n)}}}
\def\cA{{{\cal A}}}
\def\cF{{{\cal F}}}
\def\tV{\widetilde V}
\def\tW{\widetilde W}
\def\tH{\widetilde H}
\def\tE{\widetilde E}
\def\tF{\widetilde F}
\def\tA{\widetilde A}
\def\im{{{\rm i}}}
\def\tY{{{\wtd Y}}}
\def\ep{{\epsilon}}
\def\vep{{\varepsilon}}
\def\R{\rlap{\rm I}\mkern3mu{\rm R}}
\def\bD{{{\bar D}}}
\def\cD{{\cal D}}
\def\R{\rlap{\rm I}\mkern3mu{\rm R}}
\def\bD{{{\bar D}}}
\def\R{{{\Bbb R}}}
\def\C{{{\Bbb C}}}
\def\H{{{\Bbb H}}}
\def\CP{{{\Bbb C}{\Bbb P}}}
\def\RP{{{\Bbb R}{\Bbb P}}}
\def\Z{{{\Bbb Z}}}
\def\bA{{{\Bbb A}}}
\def\bB{{{\Bbb B}}}
\def\bC{{{\Bbb C}}}
\def\bD{{{\Bbb D}}}
\def\bZ{{{\Bbb Z}}}
\def\Re{{{\frak{Re}}}}
\def\Im{{{\frak{Im}}}}
\def\cosec{{\,\hbox{cosec}\,}}
\def\tX{{{\wtd X}}}

\def\hhtr{Hawking--Hunter--Taylor-Robinson\ }

\newcommand{\mitchell}{\it George P. \& Cynthia W.
Mitchell Institute for Fundamental Physics,\\
Texas A\&M University, College Station, TX 77843-4242, USA}
\newcommand{\tamphys}{\it Center for Theoretical Physics,
Texas A\&M University, College Station, TX 77843, USA}
\newcommand{\umich}{\it Michigan Center for Theoretical Physics,
University of Michigan\\ Ann Arbor, MI 48109, USA}
\newcommand{\upenn}{\it Department of Physics and Astronomy,
University of Pennsylvania\\ Philadelphia,  PA 19104, USA}
\newcommand{\SISSA}{\it  SISSA-ISAS and INFN, Sezione di Trieste\\
Via Beirut 2-4, I-34013, Trieste, Italy}

\newcommand{\ihp}{\it Institut Henri Poincar\'e\\
  11 rue Pierre et Marie Curie, F 75231 Paris Cedex 05}

\newcommand{\damtp}{\it DAMTP, Centre for Mathematical Sciences,
 Cambridge University\\ Wilberforce Road, Cambridge CB3 OWA, UK}
\newcommand{\itp}{\it Institute for Theoretical Physics, University of
California\\ Santa Barbara, CA 93106, USA}

\newcommand{\auth}{{\Large G.W. Gibbons\hoch{\sharp}, M.J. Perry\hoch{\sharp}
and C.N. Pope\hoch{\ddagger} } }

\begin{document}
\begin{flushright}
\hfill{DAMTP-2004-87\ \ \ MIFP-04-17}\\
\hfill{hep-th/0408217}
\end{flushright}


\begin{center}
{ \large {\Large\bf  The First Law of Thermodynamics for
    Kerr-Anti-de Sitter Black Holes 
\\ 
 }}

\vspace{30pt}
\auth

\vspace{30pt}
{\hoch{\sharp}\damtp}

\vspace{3pt}
{\hoch{\ddagger}\mitchell}

\vspace{30pt}

\underline{ABSTRACT}
\end{center}

   We obtain expressions for the mass and angular momenta of rotating
black holes in anti-de Sitter backgrounds in four, five and higher
dimensions.  We verify explicitly that our expressions satisfy the
first law of thermodynamics, thus allowing an unambiguous
identification of the entropy of these black holes with $\ft14$ of the
area.  We find that the associated thermodynamic potential equals the
background-subtracted Euclidean action multiplied by the temperature.
Our expressions differ from many given in the literature.  We
find that in more than four dimensions, only our expressions 
satisfy the first law of thermodynamics.  Moreover, in all dimensions
we show that our expression for the mass coincides
with that given by the conformal conserved charge introduced by
Ashtekar, Magnon and Das. We indicate the relevance of these
results to the AdS/CFT correspondence.

{\vfill\leftline{}\vfill \vskip 5pt \footnoterule 
{\footnotesize  \hoch{\ddagger} Research supported in part by DOE
grant DE-FG03-95ER40917.\vskip  -12pt}}

\pagebreak
\setcounter{page}{1}

\tableofcontents
\addtocontents{toc}{\protect\setcounter{tocdepth}{3}}
\vfill\eject

\section{Introduction}

Although of no great direct astrophysical or cosmological importance,
there has been considerable interest of late in the properties of
black holes in higher dimensions and in backgrounds with a negative
cosmological constant.  The main reason for this interest has been the
conjectured AdS/CFT correspondence, which allows statements to be made
about four-dimensional quantum field theory using solutions of the
Einstein equations with a negative cosmological constant in five
dimensions. In particular the thermodynamic properties of
five-dimensional rotating black holes in an anti-de Sitter background
give information about four-dimensional quantum field theory at
non-zero temperature in a state of rigid rotation. There are obvious
extensions to higher and lower dimensions, and to charged black holes.
The requisite five-dimensional solution, and some specialised
higher-dimensional solutions, have been available for some time
\cite{HHTR}.  In recent work \cite{gilupapo}, the most
general uncharged Kerr-de Sitter solutions in arbitrary dimension $D$
have been constructed, and also some charged solutions in $D=5$
\cite{cvlupo1,cvlupo2}.
  
    An examination of the fairly extensive literature reveals
that while everyone seems agree as to the area $A$, the temperature $T=
\kappa/( 2 \pi)$ and the angular velocities $\Omega _i$ relative to a
non-rotating frame at infinity, there is a lack of unanimity about
what precisely are the correct expressions for the total mass or
energy $E$, and the total angular momenta $J_i$, of the Kerr-anti-de Sitter
black holes.  Even more worrying, the proffered answers do not always
satisfy the first, and hence the second, law of thermodynamics. 
The principal purpose of the present paper is to rectify this situation. 
To recall the words of Eddington \cite{E}, much quoted during the first
gravitational thermodynamic revolution,

\begin{quote}
{\it The law that entropy always increases -- the second law of 
thermodynamics -- holds, I think, a supreme position among the laws of
Nature.
 If someone points out to you
that your pet theory of the universe is in disagreement with
Maxwell's 
equations, 
then so much the worse for Maxwell's equations. If it is found to
be
contradicted by observations -- well, these experimentalists do bungle
sometimes.
But if your theory is found to be against the second law of
thermodynamics
 I can give you no hope;
 there is nothing  for it but to collapse in humiliation.}
\end{quote} 

   Perhaps the most elementary approach is to follow the original
path of Christodoulou and Ruffini \cite{christ,christruffin}. That is, to
note that a particle of 4-momentum $p_\mu$ will have a conserved
energy\footnote{We use the mainly plus signature convention.}
$e=-p_\mu K^\mu$ and conserved angular momentum $j= p_\mu m^\mu$,
where $K^\mu$ is the Killing vector, timelike near infinity, associated to a
non-rotating frame at infinity, and $m^\mu$ is the angular Killing
field, with closed orbit normalised to have
parameter length $2 \pi$. The future directed null generator $l^\mu$
of the horizon of the black hole is given by
\ben
l^\mu = K^\mu + \Omega\, m^\mu\,.
\een 
If the particle crosses the event horizon, 
 since the momentum is future directed timelike or null, we shall have
\ben
l^\mu\, p_\mu  \le 0\,,
\een 
or in other words, 
\ben
-e +\Omega\, j \le 0\,.  
\een
Now energy and angular momentum of the particle presumably
increase the mass or total energy\footnote {Here we are using the notation
$dE$ in the usual sense  of thermodynamics.}  $E$ by $dE=e$
and the angular momenta by $dJ =j$,
whence
\ben
dE - \Omega dJ  \ge 0\,. \label{second}
\een

Christodoulou and Ruffini went further; they noticed that the
left-hand side of (\ref{second}) is proportional to an exact differential,   
and they were led to introduce an irreducible mass $M_{\rm irr}$ 
and to express $E$ as a function of $M_{\rm irr}$ and $J_i$.  For example, 
in the case of the four-dimensional
Kerr solution (with vanishing cosmological constant), they found
what we shall call the {\it Christodoulou-Ruffini Mass Formula} 
\ben
E= \sqrt{ M^2 _{\rm irr} + {J^2 \over 4 M^2 _{\rm irr} }  }\,.
\een

Christodoulou and Ruffini  regarded their mass formula as analogous
to the special relativistic relation for the kinetic energy 
$(E-M_{\rm irr})$ of a particle of rest mass $M_{\rm irr}$,  
\ben
E= \sqrt {M^2 _{\rm irr}  + {\bf p} ^2 }\,. 
\een

    Subsequent work revealed that
\ben
M_{\rm irr} = \sqrt {{ A \over 16 \pi}} \,, 
\een
where $A$ is the area of the event horizon,
and that (\ref{second}) may be written as
\ben
dE - \Omega _i dJ_i  = { \kappa \over 8 \pi} dA \ge 0\,. \label{secondsecond}
\een
This, and Hawking's area-increase law, led to the idea that some
multiple of $A$ should play the role of entropy $S$. The multiple, $ {1
\over 4} $,
was subsequently fixed by Hawking's discovery of the quantum radiance
of black holes. The Christodoulou-Ruffini mass formula could  then be
recognised  as giving the shape of the Gibbs surface in the three-dimensional
affine space with coordinates $(S, E, J)$. 

Mathematically, the 1-form 
\ben
dE-TdS -\Omega dJ
\een 
can be shown to define a contact structure on the five-dimensional space with
affine coordinates $(E,T,S,\Omega,J)$,  and the Gibbs surface is the
projection down to the three-dimensional space spanned by $(E,T,S)$ 
of a Legendrian
submanifold; \ie the analogue for contact geometry of
a Lagrangian submanifold of a symplectic manifold. Indeed, suppressing
the $E$ coordinate gives a symplectic four-dimensional  space  
with coordinates
 $(T,S,\Omega, J)$ and  symplectic form    
\ben
dT \wedge dS + d \Omega \wedge d J\,.
\een
The two-dimensional  surface in four-dimensional space given, say,  
by expressing $T$  and $\Omega$ as functions of $S$ and $J$ is a
Lagrangian submanifold. If the energy is obtained by minimizing
the total energy of a system with fixed entropy $S$ and angular momentum
$J$, then, as we shall remind the reader later in detail, 
it is also a mathematical triviality that the first law must
hold, which perhaps explains the forcefulness with  which Eddington
expressed himself.

All these relationships were originally derived by using
the particular form of the Kerr solution. However, a general
derivation for asymptotically flat solutions
with $\Lambda =0$  was then given, by first giving a general
proof, using Komar integrals,  of another  formula which had previously been
obtained directly from the Kerr metric functions, namely
\ben
 \ft12 E= { \kappa A \over 8  \pi }   + \Omega J \,. \label{Smarr}  
\een
Variation of (\ref{Smarr}) through a family of stationary vacuum solutions  
yields (\ref{secondsecond}), \ie the first law.
In fact rewriting (\ref{Smarr}) as
\ben
\ft12 E =  TS    + \Omega J \,. \label{Duhem}
\een 
allows us to recognise it as a generalised {\it Gibbs-Duhem relation},
arising from the first law by a simple scaling argument 
based on the elementary  fact of  dimensional analysis that 
in four spacetime dimensions, 
$E$ must be a homogeneous function, of degree $\ft12$, in $A$ and
$J$.

   We can also interpret the {\it Smarr-Gibbs-Duhem relation} as
 telling us that the thermodynamic potential
\ben
\Phi =\Phi( T, \Omega)= E-TS-\Omega J
\een 
is given simply by
\ben
\Phi = \ft12 E\,. \label{Smarr-Gibbs-Duhem}
\een

   The Smarr-Gibbs-Duhem relation 
(\ref{Smarr-Gibbs-Duhem})\footnote{In the elementary thermodynamics 
of a single-component homogeneous gas,
the energy is a homogeneous function of degree of the entropy and
volume, and the Gibbs-Duhem relation says that the Gibbs free energy
$G=E-TS+PV$ vanishes.}, if it holds, is sometimes useful in the
Euclidean approach to quantum gravity, because in the statistical
mechanics of a Gibbs ensemble of fixed temperature and angular
velocity, the partition function satisfies
\ben
Z(T,\Omega) = e^{- \beta {\Phi}}\,, 
\een 
and therefore thermodynamic potential $\Phi$ is expected to be related
to the Euclidean action $I$ quite generally by the {\it Quantum
Statistical Relation}\footnote{Note that the Quantum Statistical
Relation necessarily contains Planck's constant and is
quantum-mechanical in origin. It {\sl cannot} be deduced from
classical thermodynamics and is logically independent of classical
thermodynamics.  For this reason it seems inappropriate to refer to
(\ref{euclid}) as the Gibbs-Duhem relation, as is done by some
authors, since Gibbs died in 1903 and Duhem in 1916, before the advent
of quantum mechanics.  The adoption of (\ref{euclid}) for quantum
gravity was first proposed in \cite{gibhaw}.}
\ben \fbox{$ \displaystyle
\beta \Phi = I  \label{euclid}
$}\een  
\ie $Z\equiv e^{-\beta\Phi}=e^{-I}$. The action $I(\beta,\Omega)$ is the 
Legendre  transform of the entropy
$S=S(M,J)$, and is referred to in thermodyamics as the Massieu function,
often denoted by the symbol $\Gamma$.

  If the Smarr relation holds then
\ben
I= \ft12  E \beta. 
\een 
Note that if we have some way of calculating the thermodynamic
potential $\Phi$, say by a Euclidean path integral, then we
can calculate, or indeed define,
the entropy $S$ and angular velocities $J  $, via
\ben
S=- {\partial \Phi \over \partial T} \,,\qquad J  = -{\partial  \Phi
  \over \partial \Omega}\,. 
\een 
The first law now follows trivially by a Legendre transformation, the
mass $E$ being defined by
\ben
E= \Phi + TS + \Omega J\,. 
\een
Of course the quantities $E$ and $J$ so obtained  
need have no direct relation
to those  defined using, say, the ADM mass and angular momentum.
In the case of asymptotically-flat black holes it is well known
that at the semi-classical level these two methods agree.
If one recalls that
\ben
S= \beta {\partial I \over \partial \beta} -I,
\een
one realises that the  
perhaps puzzling possibility,  that  
 a classical Euclidean solution of the equations of motion 
can have a positive entropy, depends on the Euclidean action $I$'s
not being {\it proportional} to the imaginary time period 
 $\beta$, as it would for an ordinary static soliton for example,
but instead having a steeper dependence on $\beta$.             

   Despite these successes in the asymptotically-flat case, it is
important to recognise that there are obvious difficulties in deriving
the Smarr formula, and its variation, if the cosmological constant is
negative.  Various infinite integrals are encountered, and even the
evaluation of the Komar integrals requires care \cite{Magnon}.
Moreover, there is no longer a scaling symmetry, although it may be
re-instated to some extent by rescaling the cosmological constant.
Thus, while the extension of the above ideas to asymptotically flat
vacuum rotating black holes in higher dimensions follows in a
straightforward fashion, once the solutions are known, the extension
to asymptotically anti-de Sitter rotating black holes is, as we shall
see, not so simple.   

    The plan of the paper is as follows.  In section \ref{foursec}, we
treat four-dimensional Kerr-AdS black holes, and show that with the 
correct choice of energy and angular momentum, consistent with $SO(3,2)$ 
invariance, the first law of thermodynamics holds, and the entropy is
given by $\ft14$ of the area $A$.  We also verify that the quantum 
statistical relation (\ref{euclid}).  The section concludes with a 
detailed comparison with other expressions given in the literature
for both the energy and the angular momentum.  We show that these do
not satisfy the first law of thermodynamics.
  
   Section \ref{fivesec} is devoted to the five-dimensional case,
where we obtain analogous results that are consistent with the first
law of thermodynamics.  Again we give a comparison with various
expressions suggested in the literature, showing that they fail to
satisfy the first law.  Section \ref{consec} contains a description of
the two different approaches to the conformal boundary at infinity,
which seem to be responsible for the differences between the results
in the literature.

   Section \ref{d6sec} generalises these results to all dimensions $D\ge 6$.
We give results for the mass, angular momenta and actions for the general
Kerr-AdS metrics in all dimensions.  Some details of the calculation of
the action are relegated to Appendix A.  

   In section \ref{ashsec}, we show that in all dimensions, our
expression for the mass of the general Kerr-AdS black hole coincides
with that given by the conformal conserved charge introduced by
Ashtekar, Magnon and Das \cite{ashmag,ashdas}.

   Finally, section \ref{consec} contains our conclusions.

\section{Kerr-AdS Black Holes in Four Dimensions}\label{foursec}

\subsection{Thermodynamics in four-dimensional Kerr-AdS}

   The four-dimensional Kerr-(anti)-de Sitter metric was obtained by 
Carter \cite{carter}.  It can be written as 
\be
ds_4^2 = -\fft{\Delta}{\rho^2}\, \Big[dt - \fft{a}{\Xi}\, \sin^2\theta\, 
d\phi\Big]^2 + \fft{\rho^2\, dr^2}{\Delta} +
   \fft{\rho^2\, d\theta^2}{\Delta_\theta} + 
    \fft{\Delta_\theta\, \sin^2\theta}{\rho^2}\, \Big[ a\, dt - 
           \fft{r^2+a^2}{\Xi}\,d\phi\Big]^2\,,
\ee
where
\bea
\Delta &\equiv & (r^2 + a^2)(1 + r^2\, l^{-2}) -2m\, r\,,\qquad 
\Delta_\theta \equiv  1-a^2\, l^{-2}\, \cos^2\theta\,,\\
\rho^2 &\equiv& r^2 + a^2\, \cos^2\theta\,,\qquad 
\Xi\equiv 1 - a^2\, l^{-2}\,,
\eea
and it satisfies $R_{\mu\nu}= -3 l^{-2}\, g_{\mu\nu}$. 
    
        The outer horizon is located at the radius $r=r_+$, where $r_+$ is
the largest root of $\Delta=0$.  The area of the event horizon is
\be
A ={  4 \pi ( r_+^2 +a ^2 ) \over \Xi}\,.
\ee
The surface gravity $\kappa$ and inverse temperature $\beta$ are given by
\be
\beta =T^{-1}= \fft{2\pi}{\kappa}  = {4\pi ( r_+^2 + a^2 ) \over r_+ 
    ( 1 + a^2\, l^{-2} + 3
 r_+^2\, l^{-2}   - a ^2 r_+^{-2} ) }\,. \label{kappa}
\ee

    The angular velocity of the black hole, measured relative to a frame 
that is {\sl non-rotating at infinity}, is given by
\ben
\Omega = {a ( 1+ r_+^2 \, l^{-2} ) \over r_+^2 +a ^2}  \,.\label{angvel}
\een
Note that Hawking, Hunter and Taylor-Robinson define an angular velocity
which is measured relative to a frame rotating at infinity, by
\ben
\Omega ^\prime = { a \X \over r_+^2 + a^2 }\,,\label{angvelp}
\een
and so
\ben
\Omega -\Omega ^\prime = { a \over l^2 }\,.\label{angveldiff}
\een

  The action $I_4$ of the four-dimensional Kerr-AdS black hole turns out
to be given by
\bea
I_4 &=& \fft{\beta}{2\Xi}\, [m - r_+\, l^{-2}\, (r_+^2+a^2)] \,,\\
&=& - { \pi (r_+^2 + a ^2 )^2
 ( r_+^2 \, l^{-2} -1)  \over \Xi \, l^2 \,(3
  r_+^4 \, l^{-2}    + (
  1 +  a^2\, l^{-2} )\,  r_+^2  -a^2) } \,.
\eea
(See Appendix A for a discussion of the calculation of the action of 
the Kerr-AdS metrics in arbitrary dimension.)

The mass and angular momentum are related to the parameters $m$ and $a$
appearing in the metric; they  were
first obtained by Henneaux and Teitelboim \cite{HT}.
Their approach was a Hamiltonian one,  and they
obtained the generators of $SO(3,2)$, i.e. in modern language
the values of the moment maps for the action of $SO(3,2)$ on the
gravitational phase space, and checked their Poisson brackets.
For these solutions the Abbot-Deser masses \cite{abodes} and the 
Ashtekar-Magnon \cite{ashmag} 
masses  give the same answers. The Komar technique
of Magnon \cite{Magnon} also gives the same answers for  angular momentum
and, with care, for the mass.
Henneaux and Teitelboim find the dimensionless $SO(3,2)$ generators to be
\ben
J_{51}= {16 \pi  ml   \over \Xi^2 }\,,\qquad J_{23}= -{16 \pi ma
  \over \Xi^2 }\,.
\een

   The above considerations lead naturally to the following expressions
for the ``physical'' mass (or energy) $E$ and angular momentum $J$:
\ben 
\fbox {$\displaystyle E= {m \over \Xi^2 }\,,\qquad
J= {ma \over \Xi^2}\, \label{right} $}\label{rightej}
\een
We shall discuss different choices that can be found in the literature
later.  For now, we note that Kostelecky and Perry \cite{kosper} obtained
the generators for Kerr-Newman black holes, and that their expressions 
agree with those of Henneaux and Teitelboim in the uncharged case.
It is also worth remarking that the calculation of
$J$, either via the Komar integral or other approaches, is relatively
straightforward and unambiguous. By contrast, the calculation of $E$ is
trickier; for example, one encounters a divergence when performing a 
Komar integral, and this leads to ambiguities in performing 
a subtraction to obtain a finite result.

  It is easily verified that with the definitions (\ref{right}) of
mass and angular momentum, and with the angular velocity $\Omega$
relative to a non-rotating frame at infinity given by (\ref{angvel}),
then the first law of thermodynamics holds, namely
\ben 
dE=T\, dS+\Omega \, dJ\,,\label{de}
\een
with the entropy 
\be
S=\ft14 A\,.\label{s4}
\ee
Note that our identification of the entropy $S$ with $\ft14$ of the area
$A$ depends crucially upon the fact that the first law of
thermodynamics holds; in particular, the right-hand side of (\ref{de})
is exact if and only if the coefficient is $\ft14$.  Without this, we
could merely give a hand-waving argument based on the idea that both
$S$ and $A$ are non-decreasing functions.

It is also easily verified that these quantities satisfy the Quantum 
Statistical Relation
\be
E-T\, S -\Omega\, J = T\, I_4\,.\label{qsr1}
\ee

    It is useful to note that one can reverse the logic, and calculate
the energy $E$ (up to an additive constant) by integration of
(\ref{de}).  Since, as we remarked above, the expression for $J$ is on
``firmer ground'' than that for $E$, and of course the expressions for
$T$, $S$ and $\Omega$ are uncontroversial, this can provide a useful
way of checking the validity of a proposal for $E$.  It is, of course,
essential that the right-hand side of (\ref{de}) be an exact
differential if this procedure to determine $E$ is to make sense.
With $T$, $\Omega$, $J$ and $S$ given as in (\ref{kappa}),
(\ref{angvel}), (\ref{right}) and (\ref{s4}), it is straightforward to
verify that the right-hand side of (\ref{de}) is indeed exact, and by
integration one can then recover the expression for $E$ given in
(\ref{right}).  It is natural to choose the constant of integration so
that $E$ vanishes when the parameter $m$ vanishes, \ie for the AdS
metric.\footnote{In the literature on the Ads/CFT correspondence this
is not always done, and in five dimensions, but not in four
dimensions, $E$ contains a Casimir contribution.  This of course would
break $SO(4,2)$ invariance.}

\subsection{Comparison with other literature}

   In some of the literature, a different expression for the mass or
energy of the four-dimensional Kerr-AdS black hole has been given.  
For example, Hawking, Hunter and Taylor-Robinson give the expression
\be
E' = \fft{m}{\Xi}\label{eprime}
\ee
for the mass, while still taking $J=m\, a/\Xi^2$, as in (\ref{right}), for
the angular momentum.  They note that with this choice, the ``usual 
thermodynamic relations'' imply that the entropy is indeed given by
(\ref{s4}), as one would expect.  By this, they appear to mean that
a version of the Quantum Statistical Relation holds, namely
\be 
E' -T\, S - \Omega'\, J = T\, I_4\,.\label{qsr2}
\ee
Note that this version differs from the relation (\ref{qsr1}) that is 
satisfied by the 
mass $E=m/\Xi^2$, since there the angular velocity $\Omega$, given in
(\ref{angvel}),
measured relative to a non-rotating frame at infinity was used, whereas
in (\ref{qsr2}) it is the angular velocity $\Omega'$, given in
(\ref{angvelp}), measured 
relative to a frame rotating at infinity, that must be used.  The relation
between (\ref{qsr1}) and (\ref{qsr2}) can easily be understood by noting
from (\ref{angveldiff}), (\ref{right}) and (\ref{eprime})  that
\be
E' - \Omega'\, J = E - \Omega\, J\,.
\ee

   What is not commented upon in \cite{HHTR} is the issue of the first
law of thermodynamics.  It is easily verified that not only is it
not satisfied with these definitions,  \ie
\be
dE' \ne T\, dS + \Omega'\, dJ\,,\label{nonfl}
\ee
but the right-hand side of (\ref{nonfl}) is not even an exact
differential, and thus it cannot be integrated to give any energy
function.  In other words, at least if one adopts the
universally-accepted expression $J=m\, a/\Xi^2$ for the angular
momentum, it is impossible to satisfy the first law of thermodynamics
if the angular velocity $\Omega'$, defined in (\ref{angvelp}), is used
in (\ref{nonfl}).  Rather, one should use the angular velocity
$\Omega$, which is measured relative to a non-rotating frame at
infinity, and which is defined in (\ref{angvel}).\footnote{The
importance of measuring the angular velocity relative to a
non-rotating frame was emphasised by Caldarelli, Cognola and Klemm
\cite{calcog}.}  As we have noted, by integrating the right-hand side of
(\ref{de}) one obtains the result $E=m/\Xi^2$ that was given in
\cite{HT,kosper}.

    Caldarelli et al. \cite{calcog} agree with our formulae
(\ref{rightej}), which they obtained using the Brown-York
\cite{broyor} method.  Silva \cite{silva} agrees with the angular
momentum in (\ref{rightej}) but obtains for the energy the same
expression (\ref{eprime}) as Hawking, Hunter and Taylor-Robinson.  He
notes without comment that to get the first law to turn out correctly,
he must divide his expression for the energy by $\Xi$.

    It is interesting to note that in \cite{calcog} a so-called
``Smarr formula'' (which we prefer to call a ``Christodoulou-Ruffini
mass formula'') is derived.  It is given by
\ben
E= \sqrt{M^2_{\rm irr}\, (1 + 4 M^2_{\rm irr}\, l^{-2})^2 +
     \ft14 J^2 M_{\rm irr}^{-2} (1+ 4 M^2_{\rm irr}\, l^{-2})}\,.
\een
The failure of the Smarr-Gibbs-Duhem relation in the presence of the
cosmological constant, due to its breaking of scale invariance,  
may be seen from the fact that
\be
\ft12 E -T S -\Omega  J =\fft{(r_+^2+a^2)(a^2\, r_+^2 - a^2\, l^2 
     - 2 l^2 \,r_+^2)}{4(a^2-\ell^2)^2\, r_+}\ne 0\,.
\ee
(See \cite{rash} for a discussion of thermodynamics in a system of
gravity coupled to non-linear electrodynamics, where the absence of a
scaling symmetry also leads to a breakdown of the Smarr formula, but the
first law continues to hold.)

   It is also shown in \cite{calcog} that starting from the action $I_4$
one may calculate the thermodynamic potential $\Phi(T,\Omega)$, and
hence $S$ and $J$, which agree with (\ref{right}).  Performing the
Legendre transform they obtain the mass $E(S,J)$, which agrees with
(\ref{right}).  And finally, it is stated in \cite{calcog}  that these
expressions, once a partial Legendre transform has been made to the
Helmholtz free energy 
\ben F(T,J) =\Phi + \Omega J\,, 
\een 
give the correct expressions for the entropy $S$ and the angular
velocity $\Omega$.

\section{Kerr-AdS Black Holes in Five Dimensions}\label{fivesec}

\subsection{Thermodynamics in five-dimensional Kerr-AdS}

   The five-dimensional Kerr-(anti)-de Sitter metric was obtained
by Hawking, Hunter and Taylor-Robinson \cite{HHTR}.  It is given by
\bea
ds_5^2 &=& -\fft{\Delta}{\rho^2}\, \Big[ dt - 
   \fft{a\, \sin^2\theta}{\Xi_a}\, d\phi - \fft{b\, \cos^2\theta}{\Xi_b}\, 
d\psi\Big]^2 + \fft{\Delta_\theta\, \sin^2\theta}{\rho^2}\, 
\Big[ a\, dt -\fft{r^2+a^2}{\Xi_a}\, d\phi\Big]^2\nn\\
&&  \fft{\Delta_\theta\, \cos^2\theta}{\rho^2}\, 
\Big[ b\, dt -\fft{r^2+b^2}{\Xi_b}\, d\psi\Big]^2 +
\fft{\rho^2\, dr^2}{\Delta} + \fft{\rho^2\, d\theta^2}{\Delta_\theta} \nn\\
&& + \fft{(1+ r^2\, l^{-2})}{r^2\, \rho^2}\, 
\Big[ a\, b\, dt - \fft{b\, (r^2+a^2)\, \sin^2\theta}{\Xi_a}\, d\phi 
- \fft{a\, (r^2+b^2)\, \cos^2\theta}{\Xi_b}\, d\psi\Big]^2\,,
\label{hawkmet}
\eea
where 
\bea
\Delta &\equiv & \fft1{r^2}\, (r^2+a^2)(r^2+b^2)(1 + r^2\, l^{-2}) -2m\,,\nn\\
\Delta_\theta &\equiv& 1 -a^2\, l^{-2}\, \cos^2\theta -  
           b^2\, l^{-2}\, \sin^2\theta\,,\nn\\
\rho^2 &\equiv& r^2 + a^2\, \cos^2\theta + b^2\, \sin^2\theta\,,\nn\\
\Xi_a &\equiv& 1 - a^2\, l^{-2}\,,\qquad \Xi_b \equiv 1- b^2\, l^{-2}\,.
\eea
The metric satisfies $R_{\mu\nu}=-4 l^{-2}\, g_{\mu\nu}$.

   The outer event horizon is located at the largest positive root 
$r=r_+$ of $\Delta=0$. It has area $A$ given by
\be
A= \fft{2\pi^2\, (r_+^2+a^2)(r_+^2+b^2)}{r_+\, \Xi_a\, \Xi_b}\,.
\ee
The surface gravity $\kappa$ is given by
\be
\kappa = r_+\, (1 + r_+^2\, l^{-2})\Big( \fft1{r_+^2+ a^2} + 
        \fft1{r_+^2+ b^2}\Big) - \fft1{r_+}\,.
\ee
The Hawking temperature is $T=1/\beta=\kappa/(2\pi)$.

   The angular velocities, relative to a non-rotating frame at infinity,
are given by
\be \fbox{$\displaystyle
\Omega_a = \fft{a\, (1 + r_+^2\, l^{-2})}{r_+^2 + a^2}\,,\qquad
\Omega_b = \fft{b\, (1 +r_+^2\, l^{-2})}{r_+^2 + b^2}\label{abang}
$}\ee
The action of the five-dimensional Kerr-AdS metric is given by \cite{HHTR}
\be \fbox{$\displaystyle 
I_5 = \fft{\pi\, \beta}{4\Xi_a\, \Xi_b}\, [ 
   m - l^{-2}\, (r_+^2 + a^2)(r_+^2+b^2) ]\label{d5action}
$} 
\ee
This result has been obtained using the background subtraction method,
in which one regularises the infra-red divergence associated with the
infinite volume by subtracting the action of AdS, with a boundary at
large distance $R$ matched to the boundary of Kerr-AdS, and then sends
$R$ to infinity.  In general this procedure differs from the Brown-York 
procedure, widely used in AdS/CFT studies, in which one calculates a 
regularised action using a particular conformal choice of metric
representative on the conformal boundary.  The answer one obtains using
the Brown-York procedure depends non-trivially on the choice of 
conformal representative. (See Appendix A for a discussion of the derivation
of the Kerr-AdS action using the background subtraction method.)

   There is even less agreement in $D=5$ than there was in $D=4$ about the
correct expressions for the mass or energy $E$, and the angular momenta. 
However, as in four dimensions the evaluation of Komar integrals for the
angular momenta is unambiguous.  Again, for the mass one encounters
a divergence in the Komar integral, with all the associated possibilities 
for ambiguity in extracting a meaningful finite result by subtraction.

    The formulae for mass and angular momentum that we shall adopt are 
as follows:
\be \fbox{$\displaystyle 
E = 
\fft{\pi\, m\,  (2\Xi_a + 2\Xi_b -\Xi_a\, \Xi_b)}{4 \Xi_a^2 \, \X_b^2}\,,
\qquad 
J_a = \fft{\pi\, m\,  a}{2 \Xi_a^2\, \Xi_b}\,,\qquad
 J_b = \fft{\pi\, m\,  b}{2 \Xi_b^2\, \Xi_a}\label{ejdefs}
$} \ee
The angular momenta are those that result from Komar
integrals.  We have arrived at our formula for the mass by thermodynamic
considerations, as we shall now describe.

   The first law of thermodynamics should now take the form
\be
dE = T\, dS + \Omega_a\, dJ_a + \Omega_b\, dJ_b\,,\label{de5}
\ee
since there are now two independent angular momenta in orthogonal 
transverse 2-planes.  It is straightforward to verify that with
$J_a$ and $J_b$ given in (\ref{ejdefs}), and $T$, $\Omega_a$ and 
$\Omega_b$ as given above, then the right-hand side of (\ref{de5}) is indeed
an exact differential, if the entropy is taken to be
\be
S = \ft14 A\,.
\ee
It can thus be integrated to give an energy 
function and, with the constant of integration chosen so that $E=0$ for
the case $m=0$ of pure AdS$_5$, the energy is precisely the one given in
(\ref{ejdefs}).

   As a consistency check on our proposal in (\ref{ejdefs}) for the mass 
$E$, we can easily verify that it indeed satisfies the Quantum Statistical
Relation, which reads
\be
E - T S -\Omega_a\, J_a - \Omega_b\, J_b = T\,  I_5\,.
\ee
Alternatively, we could have started with the action $I_5$ given in 
(\ref{d5action}), expressed it as a function of $(\beta,\Omega_a,\Omega_b)$,
and then taken the thermodynamic potential to be $\Phi=I_5/\beta$.  From
this we could calculate the entropy and angular momenta.  Doing so, one 
recovers the results that $S=\ft14 A$ and that the angular momenta are
given by (\ref{ejdefs}).

\subsection{Comparison with other literature}\label{5compsec}

   Hawking, Hunter and Taylor-Robinson \cite{HHTR} gave the following
expressions for the mass and the angular momenta of the five-dimensional
Kerr-AdS black holes:
\be
E' =\fft{3\pi\, m}{4\Xi_a\, \Xi_b}\,,\qquad
J_a' = \fft{\pi\, m\, a}{2\Xi_a^2}\,,\qquad
J_b' = \fft{\pi\, m\, b}{2\Xi_b^2}\,.\label{ejjp}
\ee
They associate their angular momenta with angular velocities $\Omega_a'$ and
$\Omega_b'$ on the horizon, which are defined relative to a
frame at infinity that is rotating:
\be
\Omega_a' = \fft{a\, \Xi_a}{r_+^2 + a^2}\,,\qquad 
\Omega_b' = \fft{b\, \Xi_b}{r_+^2 + b^2}\,.\label{rotvel}
\ee
It will be seen that $E'$, $J_a'$ and $J_b'$ all differ from the
expressions (\ref{ejdefs}) that we have adopted.  The quantities $(E',
J_a', J_b')$ satisfy neither a Quantum Statistical Relation, nor the
first law of thermodynamics:
\bea
&&dE' \ne T\, dS + \Omega_a'\, dJ_a' + \Omega_b'\, dJ_b'\,,\label{nonfl5}\\
&&E' - T\, S -\Omega_a'\, J_a' - \Omega_b'\, J_b' \ne T\, I_5\,.\label{nqsr3}
\eea
It can also be verified that the right-hand side of (\ref{nonfl5}) is
not an exact differential, and thus it cannot be integrated to get any
energy function.\footnote{These statements would also be true if one 
replaced $\Omega_a'$ and $\Omega_b'$ by $\Omega_a$ and $\Omega_b$ in
(\ref{nonfl5}) and (\ref{nqsr3}).}

   It is worth remarking that although there have been various different
proposals for the mass, most of the literature on the five-dimensional
Kerr-AdS solution agrees that the angular momenta should be given by
$J_a$ and $J_b$ in (\ref{ejdefs}), rather than $J_a'$ and $J_b'$ in
(\ref{ejjp}).  If one adopts $J_a$ and $J_b$ for the angular 
momenta, but still
follows the four-dimensional philosophy of \cite{HHTR} by using $\Omega_a'$
and $\Omega_b'$ for the angular velocities, then the mass $E'$ proposed
by Hawking, Hunter and Taylor-Robinson {\sl does} satisfy a Quantum 
Statistical Relation, namely
\be
E' - T\, S -\Omega_a'\, J_a - \Omega_b'\, J_b = T\, I_5\,.
\ee
However, one again finds that the first law fails,
\be
dE' \ne T\, dS + \Omega_a'\, dJ_a + \Omega_b'\, dJ_b\,,
\label{inexact}
\ee
and again, the right-hand side is not even an exact differential.  

    Berman and Parikh \cite{bermpari} discuss the thermodynamics of
five-dimensional Kerr-AdS black holes with a single rotation
parameter, giving expressions that coincide with those of Hawking,
Hunter and Taylor-Robinson in this special case.  In fact the angular
momentum agrees with our expression in (\ref{ejdefs}) in this special
case (because, for example, if $b=0$ then $\Xi_b=1$).  However, the
mass does not agree with our expression in (\ref{ejdefs}) even in this
special case.  Their discussion of the quantum statistical relation is
similar to that of Hawking, Hunter and Taylor-Robinson, and works out
for the same reason.  However, just as in the case of Hawking, Hunter
and Taylor-Robinson, they give no discussion of the first law, and in fact
one does not obtain an exact differential on the right-hand side of 
(\ref{inexact}).

   The moral to be extracted from the above seems to be that one should
always use the angular velocities measured relative to a non-rotating 
frame at infinity when discussing the thermodynamics of rotating 
black holes.

   In a subsequent paper, Hawking and Reall \cite{hawkreal} give
expressions for the mass and angular momenta, for which they refer to
\cite{HHTR}.  However, although the mass coincides with the expression
appearing in \cite{HHTR}, and reproduced as $E'$ in (\ref{eprime}),
the angular momenta differ from those appearing in \cite{HHTR} by a
denominator factor of $(1+ r_+^2\, l^{-2})$.

   Another proposal for the mass of the five-dimensional Kerr-AdS 
black hole is given by Awad and Johnson \cite{awad}.  Based on the 
application of the Brown-York procedure, they obtain a mass
\be
E'' = \fft{\pi\, l^2}{96 \Xi_a\, \Xi_b}\, (7 \Xi_a\, \Xi_b + 
\Xi_a^2 + \Xi_b^2 + 72 m\, l^{-2})\,,\label{e5pp}
\ee
which does not vanish in AdS$_5$ spacetime ($m=0$).  This reduces in
the non-rotating case to the expression obtained by Balasubramanian
and Kraus \cite{balakrau} 
\ben { 3 \pi l^2 \over 32} + { 3 \pi m \over 4}
\,.  
\een 
In the AdS/CFT correspondence, the first term is attributed
to the Casimir energy of the conformal field theory on the boundary.

   Awad and Johnson also present an expression for the action of the
five-dimensional Kerr-AdS black hole,
\bea
I_5'' &=& -{ \pi \beta l^2 \over 96 \X_a\X_b} \Bigl [ 12 r_+^2\, l^{-2}\, 
( 1-\X_a -\X_b) + \X^2_a + \X^2_b + \X_a \, \X_b + 12 r^4_+ \, l^{-4}\nn\\
&& \qquad\qquad -
2 (a^4 +b^4)\, l^{-4}
  - 4  a^2 b^2 \, l^{-4}\, (3 r_+^{-2}  l^{2 }- 1 )
 -12  \Bigr ]\,. \label{Awad}
\eea
This is obtained by a boundary counterterm subtraction procedure.  Not
surprisingly, this expression differs from the canonical expression
(\ref{d5action}) for the action.

   One can easily verify that indeed, as stated in \cite{awad}, the mass
$E''$ given in (\ref{e5pp}) rather non-trivially satisfies a Quantum
Statistical Relation, namely
\be
E'' - T\, S - \Omega'\, J_a - \Omega_b'\, J_b = T\, I_5''\,.
\ee
It should be noted that the generally-accepted angular momenta $J_a$ 
and $J_b$ defined in (\ref{ejdefs}) are being employed here, but the
angular velocities are the ones given in (\ref{rotvel}), which are 
defined with respect to a frame rotating at infinity.  The
first law of thermodynamics is not investigated in \cite{awad}, but
it is easy to see that it is not satisfied;
\be
dE'' \ne T\, dS - \Omega_a'\, dJ_a - \Omega_b'\, dJ_b\,.\label{eefail}
\ee
Indeed, as we already observed, the right-hand side of this expression
is not an exact differential, and so one could not achieve an 
equality in (\ref{eefail}) for {\sl any} choice of mass $E''$.

    There is a further puzzling feature of Awad and Johnson's action
$I=I_5''$.  If one considers it as a function of the inverse
temperature $\beta$, and angular velocities $(\Omega_a,\Omega_b)$ (or
$(\Omega_a',\Omega_b')$), and calculates the energy $E$, 
entropy $S$ and angular
momenta $(J_a,J_b)$ assuming that the quantum statistical relation
holds,
\bea
E &=& \fft{\del I}{\del\beta} - \fft{\Omega_a}{\beta}\, 
  \fft{\del I}{\del \Omega_a} - \fft{\Omega_b}{\beta}\, 
   \fft{\del I}{\del \Omega_b}\,,\nn\\
S &=& \beta\, \fft{\del I}{\del\beta} - I\,,\qquad
J_a = -\fft1{\beta}\, \fft{\del I}{\del\Omega_a}\,,\qquad
J_b = -\fft1{\beta}\, \fft{\del I}{\del\Omega_b}\,,\label{ejdiffs}
\eea
one does not obtain their expression (\ref{e5pp}) for the energy,
their (correct) expression $S=\ft14 A$ for the entropy, or their
(correct) expression for the angular momenta $(J_a,J_b)$.  Thus it
seems that not only does the thermodynamics of the boundary system
that they consider differ from the thermodynamics of the black holes
in the bulk, but integrating the boundary stress tensor \`a la
Brown-York, as they do, does not give the same expressions for the
energy and angular momenta as one would obtain directly by
interpreting the action as a thermodynamic potential.  In any event,
it is clear that unlike the expression $I_5$ for the action, given in
(\ref{d5action}), their action $I_5''$, given in (\ref{Awad}), cannot
be used to obtain the thermodynamic potential of the bulk black hole.
We shall comment on this further in the next subsection.

    It is perhaps worth emphasising that the differences one
encounters between the various results in the literature typically
involve terms depending upon the dimensionless ratios of the rotation
parameters divided by the AdS radius $l$ (for example, via the
quantities $\Xi_a$ and $\Xi_b$), which become irrelevant in the limit
where the cosmological constant vanishes.  Likewise, the differences
between the angular velocities (\ref{abang}) and (\ref{rotvel})
disappear in the zero cosmological constant limit.  

\subsection{Conformal infinity in the Hawking-Hunter-Taylor-Robinson metrics}
\label{confsec}

   In this subsection, we discuss conformal infinity in the 
\hhtr metrics, emphasising in particular that different choices of 
conformal factor give rise to different metrics on the conformal boundary.

   We begin by applying the transformation given in \cite{HHTR} that
converts to asymptotically ``canonical'' AdS coordinates.  In our notation,
the new coordinates will be called $(y,\hat\theta, \hat\phi,\hat\psi)$,
related to the original coordinates $(r,\theta,\phi,\psi)$ of (\ref{hawkmet}) 
by
\bea
\Sigma_a\, y^2\, \sin^2\hat\theta &=& (r^2+a^2)\, \sin^2\theta\,,\\
\Sigma_b\, y^2\, \cos^2\hat\theta &=& (r^2+b^2)\, \cos^2\theta\,\\
\hat\phi &=& \phi +  a\, l^{-2}\, t\,,\\
\hat\psi &=& \psi + b\, l^{-2}\, t\,.
\eea
We find that the metric has the following asymptotic form, in
terms of the new coordinates:
\bea
ds^2 &=& -(1+ y^2\, l^{-2})\, dt^2 + \fft{dy^2}{1 + y^2\, l^{-2} -
\fft{2m}{\Delta_{\hat\theta}^2\, y^2}} + y^2\, d\hat\Omega_3^2\nn\\
&&
 + \fft{2m}{\Delta_{\hat\theta}^3\, y^2}\, (dt-
        a\, \sin^2\hat\theta\, d\hat\phi -
   b\, \cos^2\hat\theta\, d\hat\psi)^2 + \cdots\,,
\label{metric2}
\eea
where
\bea
\Delta_{\hat\theta} &\equiv& 1 - a^2\, l^{-2}\, \sin^2\hat\theta - 
 b^2\, l^{-2}\,  
   \cos^2\hat\theta\,,\\
d\hat\Omega_3^2 &\equiv & d\hat\theta^2 + \sin^2\hat\theta\, d\hat\phi^2 +
    \cos^2\hat\theta\, d\hat\psi^2\,.
\eea
   
   We recall that a bulk spacetime $\{X,g\}$ with conformal boundary
$\{\del X, \bar h\}$ admits a conformal compactification $\{\bar X,
\bar g\}$ if $\bar X= X\sqcup \partial X$ is the closure of $X$ and
the metric $\bar g$ extends smoothly to $\bar X$ with $\bar g = \Omega^2\,
g$ for some function $\Omega$ with $\Omega>0$ in $X$ and $\Omega=0$ on
$\del X$, with $d\Omega\ne0$ on $\del X$.  Since $\Omega$ is determined only 
up to a factor, $\Omega\rightarrow f\, \Omega$, where the function
$f$ is non-zero on $\del X$, the metric $\bar g$ on $\bar X$ and its 
restriction $\bar h=\bar g|_{\del X}$ are defined only up to a conformal
factor.  The conformal equivalence class $\{\del \bar X,\bar h\}$ is 
called the conformal boundary of $X$.   

   In our case, we may take
\be
\Omega= \fft{l}{y}\,,
\ee
so that $\del X$ is given by $y=\infty$ and a short calculation shows
that $\bar h$ for this choice of conformal factor is given by
\be
d\bar s^2 = - dt^2 + l^2\, d\Omega_3^2\,.\label{einststat}
\ee
This is the standard metric on the Einstein static universe, or, equivalently,
on the conformal compactification of four-dimensional Minkowski 
spacetime.
 
    Working in the  $(y,\hat\theta, \hat\phi,\hat\psi)$ coordinates is
rather clumsy, and we may instead use the Boyer-Lindquist coordinates
$(r,\theta,\phi,\psi)$ of equation (\ref{hawkmet}).  In this case it is
natural to choose
\be
\Omega=\fft{l}{r}\,,
\ee
for which one finds that the metric on the conformal boundary is given by
\be
d{{\bar {s'}}}^2 = -dt^2 + \fft{2a\, \sin^2\theta}{\Xi_a}\, dt\, d\phi
+ \fft{2b\, \cos^2\theta}{\Xi_b}\, dt\, d\psi + 
  \fft{l^2\, d\theta^2}{\Delta_\theta} + \fft{l^2\, \sin^2\theta}{\Xi_a}\,
d\phi^2 + \fft{l^2\, \cos^2\theta}{\Xi_b}\, d\psi^2\,.
\label{dogsb}
\ee
Using the coordinate transformations
\be
\tan^2\theta = \fft{\Xi_a}{\Xi_b}\, \tan^2\hat\theta\,,\qquad
\phi = \Phi -a\, l^{-2}\, t\,,\qquad 
\psi =\Psi - b\, l^{-2}\, t\,,
\ee
we find that the two boundary metrics (\ref{einststat}) and (\ref{dogsb})
are indeed conformally related, with
\be
d{\bar {s'}}^2 = \fft1{\Delta_{\hat \theta}}\, d\bar s^2\,.
\ee
In fact, this conformal factor relating the two boundary metrics is
essentially given by $\Delta_{\hat\theta}^{-1} = y^2/r^2$.

    An important point  is that the static boundary metric $d{\bar {s'}}^2$ 
has $g_{tt}= - 1/\Delta_{\hat\theta}$. This means that in thermal 
equilibrium the temperature will be $\hat\theta$ dependent, because
of Tolman's well-known red-shifting law, $T\, \sqrt{-g_{tt}} =$ constant
\cite{tolman}.  In the present case, this 
tells us that the local temperature $T=T(\hat\theta)$ is given by
\be
T(\hat\theta) = T_0\, \sqrt{\Delta_{\hat\theta}}\,,
\ee
where $T_0$ is a constant.  This spatial dependence of the equilibrium
temperature may be responsible for the discrepancies found in section
\ref{5compsec} when we compared the energy and angular momenta given
by Awad and Johnson with the expressions that follow from
differentiating their action, as in (\ref{ejdiffs}).

   It is also important to bear in mind, as emphasised by Skenderis
\cite{skend}, that if one uses the boundary stress tensor method to
obtain actions, energies and angular momenta, these will in general
depend upon the choice of conformal representative for the boundary
metric, if the spacetime dimension is odd.  This is because in odd
spacetime dimensions, the (even-dimensional) boundary is subject to
conformal anomalies.  In particular, calculations carried out using
the conformal factor $\Omega=l/r$, which may seem natural and
convenient in the Boyer-Lindquist coordinates $(r,\theta, \phi,\psi)$,
will give different answers from those carried out using the conformal
factor $\Omega=l/y$ that is natural in the $(y,\hat\theta,
\hat\phi,\hat\psi)$ coordinates.  It may be that this accounts for
some of the differences between the formulae of Awad and Johnson
\cite{awad} (who use $\Omega=l/r$) and those of other authors.  As
Ashtekar and Das have emphasised \cite{ashdas}, there is no way of
reconciling the $SO(4,2)$ anti-de Sitter symmetry with an energy, such
as that of \cite{awad,balakrau}, that is non-vanishing in pure AdS
spacetime. We ourselves have not used the boundary stress tensor
procedure, and have endeavoured to maintain bulk diffeomorphism
invariance throughout.

\section{Kerr-AdS Black Holes in $D\ge 6$ Dimensions}\label{d6sec}

\subsection{Thermodynamics in $D\ge 6$ Kerr-AdS}

    The general Kerr-de Sitter metrics in arbitrary dimension $D$
were obtained in \cite{gilupapo}.  They have $N\equiv [(D-1)/2]$
independent rotation parameters $a_i$ in $N$ orthogonal 2-planes. 
We have $D=2N+1$ when $D$ is odd, and $D=2N+2$ when $D$ is even. Defining
$\ep\equiv (D-1)$ mod 2, so that $D=2N+1+\ep$, the metrics can be 
described by introducing $N$
azimuthal angles $\phi_i$, and $(N+\ep)$ ``direction cosines'' $\mu_i$
obeying the constraint
\be
\sum_{i=1}^{N+\ep} \mu_i^2 =1\,.\label{muconstraint}
\ee
In Boyer-Linquist coordinates, the metrics are given by 
\cite{gilupapo}
\bea
ds^2 &=& - W\, (1 + r^2\, l^{-2} )\, d\tau^2
 + \fft{2m}{U}\Bigl(W\,d\tau
 - \sum_{i=1}^N \fft{a_i\, \mu_i^2\, d\varphi_i}
  {\Xi_i }\Bigr)^2
 + \sum_{i=1}^N \fft{r^2 + a_i^2}{\Xi_i}\,\mu_i^2\,
    d\varphi_i^2 \nn\\
&&
 + \fft{U\, dr^2}{V-2m}
 + \sum_{i=1}^{N+\ep} \fft{r^2 + a_i^2}{\Xi_i}\, d\mu_i^2
 - \fft{l^{-2}}{W\, (1 + r^2\, l^{-2})}
    \Bigl( \sum_{i=1}^{N+\ep} \fft{r^2 + a_i^2}{\Xi_i}
    \, \mu_i\, d\mu_i\Bigr)^2 \,,\label{bl}
\eea
where
\bea
W &\equiv& \sum_{i=1}^{N+\ep} \fft{\mu_i^2}{\Xi_i}\,,\qquad
U \equiv  r^{\ep}\, \sum_{i=1}^{N+\ep} \fft{\mu_i^2}{r^2 + a_i^2}\,
\prod_{j=1}^N (r^2 + a_j^2)\,,\label{uwline}\\
V &\equiv& r^{\ep-2}\, (1 +r^2\, l^{-2})\,
   \prod_{i=1}^N (r^2 + a_i^2)\,,\qquad \Xi_i\equiv 1 - a_i^2\, l^{-2}\,.
\label{uvw}
\eea
They satisfy $R_{\mu\nu}=-(D-1)\, l^{-2}\, g_{\mu\nu}$.  

   The outer horizon is located at $r=r_+$, where $r_+$ is the largest
root of $V(r)-2m=0$.   The surface gravity $\kappa$ and the area $A$ of
the event horizon are given by \cite{gilupapo} 
\bea
\hbox{$D=$ odd}:&& \kappa = r_+\, (1 +r_+^2\, l^{-2})\, 
  \sum_i\fft1{r_+^2 + a_i^2} -\fft1{r_+}\,,\\
&& A = \fft{{\cal A}_{D-2}}{r_+}\, \prod_i \fft{r_+^2 + a_i^2}{\Xi_i}\,,\\
\hbox{$D=$ even}:&& \kappa = r_+\, (1 + r_+^2\, l^{-2})\, 
  \sum_i\fft1{r_+^2 + a_i^2} -\fft{1- r_+^2\, l^{-2}}{2r_+}\,,\\
&& A =  {\cal A}_{D-2} \, \prod_i \fft{r_+^2 + a_i^2}{\Xi_i}\,,
\eea
where ${\cal A}_{D-2}$ is the volume of the unit $(D-2)$-sphere:
\be
{\cal A}_{D-2}  = \fft{2 \pi^{(D-1)/2}}{\Gamma[(D-1)/2]}\,.
\label{spherevol}
\ee
The Hawking temperature is then given by $T=\kappa/(2\pi)$.
The angular velocities, measured relative to a frame that is non-rotating
at infinity, are given by
\be
\Omega_i = \fft{(1 +r_+^2\, l^{-2})\, a_i}{r_+^2 + a_i^2}\,.
\ee

   By evaluating Komar integrals, 
\be
J_i \sim \int_{S^{D-2}} {*dK_i}
\ee
where 
$K_i= \fft{\del}{\del\phi^i}$ are the angular Killing vectors, one 
one can see that the angular
momenta should be given in terms of the rotation parameters by
\be
 \fbox{$\displaystyle
J_i = \fft{m\, a_i\, {\cal A}_{D-2}}{4\pi\, \Xi_i\,(\prod_j \Xi_j)}
\label{jdef} $}\label{djdefs}
\ee
These expressions agree with those in (\ref{right}) and (\ref{ejdefs}) in
four and five dimensions.  We have explicitly evaluated the Komar integrals
in dimensions $D\le 7$.

    We now consider the first law of thermodynamics, which will read
\be
dE = T\, dS + 
\sum_i \Omega_i\, dJ_i\,.\label{firstlaw}
\ee
It can be verified that the right-hand side is indeed an exact differential,
if the expressions (\ref{jdef}) are used for the angular momenta, and if the
entropy is taken to be
\be
S=\ft14 A\,.  
\ee
This provides a non-trivial check of the correctness of
(\ref{jdef}).  We can then integrate (\ref{firstlaw}), thereby
allowing us to learn the mass $E$ of the black hole.  We find
\bea
\hbox{$D$= odd}: &&   \fbox{$\displaystyle E = 
           \fft{m\, {\cal A}_{D-2}}{4\pi\, (\prod_j \Xi_j)}\, 
\Big( \sum_{i=1}^N \fft1{\Xi_i} - \fft12\Big)\label{massodd} $}
\label{dmassdefsodd}\\
\hbox{$D$= even}: &&  \fbox{$\displaystyle E = 
   \fft{m\, {\cal A}_{D-2}}{4\pi\, (\prod_j \Xi_j)}\,
\sum_{i=1}^N \fft1{\Xi_i}\label{masseven} $}\label{dmassdefseven}
\eea
It is easily seen that in the cases $D=4$ and $D=5$, these expressions
reduce to the ones given in (\ref{right}) and (\ref{ejdefs}) respectively.
In section \ref{ashsec}, we shall calculate the masses of the general
rotating AdS black holes using the 
Ashtekar-Magnon-Das conformal mass definition, and show that they are
in agreement with our expressions (\ref{dmassdefsodd})
and (\ref{dmassdefseven}).

  By applying the standard background-subtraction procedure for
calculating the Euclidean action $I_D$ of the $D$-dimensional Kerr-AdS
metric, described in detail in Appendix A, we find that it is given by
\bea
\hbox{$D$= odd}: &&  \fbox{$\displaystyle 
                       I_D = \fft{\beta\, {\cal A}_{D-2}}{
   8\pi\, (\prod_j \Xi_j)}\,\Big( m- l^{-2}\, \prod_{i=1}^N (r_+^2 + a_i^2) 
\Big)
\label{actionodd} $}
        \\
\hbox{$D$= even}: &&  \fbox{$\displaystyle 
                  I_D = \fft{{\beta\, \cal A}_{D-2}}{
   8\pi\, (\prod_j \Xi_j)}\,\Big( m -r_+\,l^{-2}\,  
   \prod_{j=i}^N (r_+^2 + a_i^2) 
\Big)\label{actioneven} $}
\eea
It is straightforward to verify that with our results (\ref{djdefs}), 
(\ref{dmassdefsodd}) and (\ref{dmassdefseven}) for the angular 
momenta and masses of the Kerr-AdS black holes, the 
Quantum Statistical Relation 
\be
E -T\, S - \sum_i \Omega_i \, J_i = T\,  I_D
\ee
holds.

\subsection{Comparison with other literature}

   No very complete comparisons with previous results can be made in 
dimensions $D\ge 6$, since the general rotating Kerr-AdS solutions
were not available until recently.  The metrics {\sl were} obtained
in \cite{HHTR} for
the special case where all except one rotation parameter vanishes, and 
so some comparisons in these restricted cases are possible.

   Awad and Johnson \cite{awad} give expressions in $D=6$ for the mass,
angular momentum and action when (to take a concrete choice for the
specialisation) $a_1\ne0$, $a_2=0$.  They give
\be
E' = \fft{4\pi\, m}{3\Xi_1}\,,\qquad J_1 = \fft{2\pi\, m\, a_1}{3\Xi_1^2}\,,
\ee
and an expression for the action $I_6$ that is easily verified to be the
same as the specialisation of (\ref{actioneven}) to $D=6$, $a_2=0$ (after
inserting a factor of $\ft13$ in their expression whose absence is
presumably due to a typographical error.).  Their
expression for $J_1$ also agrees with our general result (\ref{jdef}),
under the specialisation $a_2=0$.  Their expression $E'$ for the mass, 
however, disagrees with the specialisation of (\ref{masseven}) to $D=6$, 
$a_2=0$, which would give
\be
E = \fft{2\pi\, m}{3\Xi_1}\, \Big(1 + \fft1{\Xi_1}\Big)\,.
\ee
As observed in \cite{awad}, their expression for $E'$ satisfies the
Quantum Statistical Relation
\be
E' - T\, S - \Omega_1'\, J_1 = T\, I_6\,,
\ee
where the angular velocity $\Omega_1' = a_1\, \Xi_1/(r_+^2 +
a_1^2)$ is defined with respect to a frame that is rotating at
infinity.  However, as usual when such an angular velocity is employed,
we find that the first law is not satisfied,
\be
dE' \ne T\, dS + \Omega_1'\, dJ_1\,,
\ee
and furthermore the right-hand side is not an exact differential.

   Awad and Johnson have also studied Kerr-AdS black holes in $D=7$, in
the special case where two of the three rotation parameters vanish, say
$a_2=a_3=0$.  They give the mass and angular momentum as
\be
E'= -\fft{\pi^2\, l^4\, (a_1^6\, l^{-6} + 5 a_1^4\, l^{-4} + 50 \Xi_1 -
          800m\, l^{-4})}{1280 \Xi_1}\,,\qquad
J_1 = \fft{\pi^2\, m\, a_1}{4\Xi_1^2}\,,
\ee
and an expression for the action $I'_7$ where, as in $D=5$, they
employ a counterterm subtraction scheme.  Their expression for the
angular momentum agrees with our general result (\ref{jdef}), specialised
to $D=7$ and $a_2=a_3=0$.  However, their expressions for 
the mass and the action disagree with ours.  They again find that
their expressions satisfy a Quantum Statistical Relation,
\be
E' - T\, S - \Omega_1'\, J_1 = T\, I_7'\,.
\ee
Although they do not address the issue, it is easy to see that
\be
dE' \ne T\, dS + \Omega_1'\, dJ_1\,,\label{7nonfl}
\ee
and so the first law is not satisfied, and furthermore 
the right-hand side of (\ref{7nonfl}) is not an exact differential. 

\section{Thermodynamic Energy and the Conformal Mass}\label{ashsec}

   Astekar and Magnon \cite{ashmag} and Ashtekar and Das \cite{ashdas}
have given a conformal definition of a conserved quantity $Q[K]$
associated to an asymptotic Killing field $K$ in an aymptotically
anti-de Sitter $D$-dimensional spacetime. The expression for $Q[K]$ is
linear in $K$ and involves an integral of certain components of the
Weyl tensor over a codimension-2 sphere lying on the conformal
boundary.  Specifically, if $\bar C^\mu{}_{\nu\rho\sigma}$ is the Weyl
tensor of the conformally rescaled metric $\bar g_{\mu\nu} =
\Omega^2\, g_{\mu\nu}$, $\bar n_\mu = \del_\mu\, \Omega$ and
\be
\bar{\cal E}^\mu{}_{\nu}= l^2 \Omega^{D-3}\, \bar n^\rho \, \bar n^\sigma\, 
    \bar C^\mu{}_{\rho\nu\sigma}
\ee
is the electric part of the Weyl tensor on the conformal boundary, then
$Q[K]$ is given by\footnote{Our sign for $Q[K]$ differs from that in 
\cite{ashdas}; we use orientation conventions such that $d\bar\Sigma_t$
is positive.}
\be
Q[K]= \fft{l}{8\pi\, (D-3)}\, \oint_{\Sigma}\, 
       \bar{\cal E}^\mu{}_{\nu}\, K^\nu\, d\bar\Sigma_\mu\,,\label{conffor}
\ee
where $d\bar\Sigma_\mu$ is the area element of the $(D-2)$-sphere section of
the conformal boundary.

   If $K_1$
and $K_2$ are two asymptotic Killing vector fields, then $a\, K_1 + 
b\, K_2$ is also an asymptotic Killikng vector field, and clearly 
\be
Q[c_1\, K_1 + c_2\, K_2]= c_1\, Q[K_1] + c_2\, Q[K_2]\,,
\ee
where $c_1$ and $c_2$ are constants.  Theses charges were evaluated by
Das and Mann \cite{dasman} for the angular momentum and energy
associated to the Killing fields $\del/\del\phi$ and $\del/\del t$ in
Boyer-Lindquist coordinates, in the case of the Kerr-AdS solutions
with only one non-zero rotation parameter, in $4\le D\le 9$ spacetime
dimensions.  They compared their results with those obtained using the
boundary stress-tensor (counterterm) on the spheroidal surfaces
$r=\,$constant. The angular momenta agree with those coming from the
boundary stress tensor in all dimensions, and they agree with those
given in this paper (after specialising our results to a single
non-vanishing rotation parameter).  They obtained agreement for the
energy in even spacetime dimensions, but their results disagreed in
odd spacetime dimensions. The conformal energies given in
\cite{dasman} also differ from our expressions (after specialising our
results to a single non-vanishing rotation parameter).

   Das and Man attributed the discrepancy between their conformal
energy and the counterterm expression to a Casimir energy.
However, they calculated the energy using the Killing field 
$K\equiv \del/\del t$, which is rotating even at
infinity.  The non-rotating timelike Killing field is 
\be
\fft{\del}{\del t} +
\fft{a}{l^2}\, \fft{\del}{\del\phi}\,.\label{nonrotk}
\ee
Thus the conformal energy with respect to it is
\be
Q[\del_t + a\, l^{-2}  \, \del_\phi ]
  = Q[\del_t] + \fft{a}{l^2}\, Q[\del_\phi]\,,
\ee
where $\del_t\equiv \del/\del t$ and $\del_\phi\equiv \del/\del \phi$.
In all cases $Q[\del_\phi]=J$, and so
\be
Q[\del_t + a\, l^{-2}  \, \del_\phi]= Q[\del_t] 
 + \fft{a}{l^2}\, J\,.\label{qj1}
\ee
If one uses the values for $Q[\del_t]$ provided in \cite{dasman}, 
one in fact finds that 
\be
\fbox{$\displaystyle
Q[\del_t + a\, l^{-2}  \, \del_\phi ]= E \,,
$}
\ee
where $E$ is our expression for the thermodynamic mass, specialised
to the case of a single non-zero rotation parameter.  

    We shall now generalise this result to the case with arbitrary
non-zero rotation parameters.  The conformal mass, which should be
calculated using the timelike Killing vector that is {\it non-rotating} at
infinity, is given by
\be M_{\rm conf}\equiv Q[\del_t + \sum_i a_i\, l^{-2} \,
\del_{\phi_i}]= Q[\del_t] + \sum_{i=1}^N\, \fft{a_i}{l^2}\,
J_i\,.\label{qj2} \ee
To evaluate $Q[\del_t]$, we note that the leading-order term in the 
expression  for the coordinate component
$C^t{}_{rtr}$ of the Weyl tensor of the physical Kerr-AdS metric $g_{\mu\nu}$
given in (\ref{bl}) is 
\be
C^t{}_{rtr} = \fft{m\, (D-2)(D-3)\, l^2}{r^{D+1}} +\cdots\label{c0101}
\ee
at large distance.  The electric component $\bar{\cal E}^t{}_t$, which is
defined on the conformal boundary, is therefore given by 
\be
\bar{\cal E}^t{}_t = m\, l^{1-D}\, (D-2)(D-3)\,.\label{e00}
\ee
 From the expression (\ref{detexp}) for the determinant of $g_{\mu\nu}$, 
we see that the volume element $d\bar\Sigma_t$ of the spacelike hypersurface
$t=\,$constant lying in the boundary is given by
\be
d\bar\Sigma_t = \fft{l^{D-2}}{\prod_j \Xi_j}\, d\Omega_{D-2}\,,
\ee
where $d\Omega_{D-2}$ is the volume element of the unit $(D-2)$-sphere.
Substituting into (\ref{conffor}) with $K=\del/\del t$, we therefore find
that $Q[\del_t]$  for the general Kerr-AdS metric in $D$ dimensions is
given by
\be
\fbox{$\displaystyle
Q[\del_t] = 
 \fft{m\, (D-2)\, 
          {\cal A}_{D-2}}{8\pi\, (\prod_j \Xi_j)}\,.
$}
\ee
Thus from (\ref{qj2}) the conformal mass is
given by
\be
M_{\rm conf}=  \fft{m\, (D-2)\, 
          {\cal A}_{D-2}}{8\pi\, (\prod_j \Xi_j)} + 
     \sum_{i=1}^N\, \fft{a_i}{l^2}\, J_i\,.
\ee
Using (\ref{djdefs}), we therefore find that
\be
\fbox{$\displaystyle
M_{\rm conf} = E\,, 
$}
\ee
where $E$ is given in (\ref{dmassdefsodd}) and (\ref{dmassdefseven}).
In other words, we have shown that our expression for the
thermodynamic mass $E$ is indeed equal to the Ashtekar-Magnon-Das
conformal mass $M_{\rm conf}$ evaluated using the
non-rotating asymptotic timelike Killing field
\be
\fft{\del}{\del t} + \sum_{i=1}^N \fft{a_i}{l^2}\,\fft{\del}{\del \phi_i}\,.
\ee

    It is worth remarking that a great advantage of the
Ashtekar-Magnon-Das definition of mass in an asymptotically AdS
background is that the integral (\ref{conffor}) directly gives a
finite result, unlike the divergent expression one obtains using the
Komar mass formula.  Thus all the amiguities that plague the Komar
prescription are avoided in the Ashtekar-Magnon-Das approach.

\section{Conclusions}\label{consec}

    In this paper, we have discussed the thermodynamics of a rotating
black hole in a background anti-de Sitter spacetime.\footnote{See
\cite{balimcsawe} for a recent discussion of the thermodynamics of
non-rotating charged AdS black holes.}  Our work has been concerned
entirely with the bulk theory, and we have maintained bulk
diffeomorphism invariance throughout.  In particular, we have
maintained the full $SO(D-1,2)$ invariance.  We have given expressions
for the masses, angular momenta, and actions of the Kerr-AdS black
hole solutions in all dimensions $D$.  The actions were calculated
using the background subtraction method.  We find, in contrast to some
earlier work, a complete consistency between our answers and the
first-law of thermodynamics. In particular, as a result we confirm
that the entropy of the black hole is given by $\ft14$ of the area, a
result that is frequently assumed, but seldom established.  We also
confirm that the bulk action $I$ and the the thermodynamic potential
$\Phi$ are related at temperature $T$ by the Quantum Statistical
Relation
\be
Z= e^{-\Phi/T} =e^{-I}\,.
\ee
We have also shown that our expression for the mass of a general
Kerr-AdS black hole in $D$ dimensions coincides with the coincides
with that given by the conformal conserved charge introduced by
Ashtekar, Magnon and Das, provided one uses the timelike Killing
vector field that is {\sl non-rotating} at infinity.

    Some of our results differ from those found in the literature, and
we have tried to explain why this is the case.  Much of the recent
work on Kerr-AdS black holes was motivated by the conjectured AdS/CFT
correspondence.  In this context, in odd bulk spacetime dimensions
there are conformal anomalies in the boundary theory, which break the
bulk $SO(D-1,2)$ invariance, and moreover are dependent upon the
choice of representative of the conformal structure on the conformal
boundary.

   Hawking, Hunter and Taylor-Robinson \cite{HHTR} suggested that the
CFT dual of a Kerr-AdS black hole is a conformal field theory that is
rotating in a background Einstein static universe.  A comparison of
their partition function with that of such a rotating gas, in the case
of no interactions, revealed the same type of singularities as the
rotation angular velocities tended to their maximum values, \ie as
$\Xi_a$ or $\Xi_b$ tended to zero.  Follow-up studies by Hawking and
Reall \cite{hawkreal}, Berman and Parikh \cite{bermpari}, and
Landsteiner and Lopez \cite{lanlop}, showed that for fixed angular
velocities but high temperatures, the thermodynamic potentials agreed
at leading order (modulo the standard factor of $\ft34$), but they
differed at sub-leading orders.  Thus there appears to be a
satisfactory agreement between the bulk and boundary theories, in the
case that the metric on the boundary is the standard metric on the
Einstein static universe.  In the work of Awad and Johnson, the
boundary metric was taken to be conformal to the standard Einstein
static universe, and hence their results differ. It would be an
interesting project to investigate in more detail the relationship
between the bulk theory and these two boundary theories.

\section*{Acknowledgments}

   We are grateful to Abhay Ashtekar for drawing our attention to
reference  \cite{dasman}, and for helpful discussions.
   G.W.G. and C.N.P. thank Bo{\u g}azi{\c c}i University, Istanbul, for 
hospitality during the course of this work. C.N.P. is also
grateful to the Relativity Group in DAMTP, Cambridge and the CERN
Theory Division for hospitality.  Many of the calculations in this
paper have been performed with the aid of Mathematica.

\section*{Note Added}

   It has recently been shown in \cite{derkat} that our expressions for the 
masses and angular momenta of the rotating AdS black holes can also be 
derived from the superpotential of Katz, Bi\v c\' ak and Lynden-Bell 
\cite{kabily}.

\bigskip\bigskip

\newpage

\appendix
\bigskip\bigskip
\centerline{\Large {\bf Appendix}}
\bigskip

\section{Euclidean Action of the Kerr-AdS Metrics}

   The Euclidean action $I_D$ of the $D$-dimensional Kerr-AdS metric can 
be calculated as follows.  The metric satisfies the Einstein 
equation $R_{\mu\nu} = -(D-1)\, 
  l^{-2}\, g_{\mu\nu}$, which can be derived from the action
\be
I'_D = -\fft1{16\pi}\, \int\sqrt{-g}\, 
  [R + (D-1)(D-2)\, l^{-2}]\, d^Dx\,. 
\ee
After Euclideanisation, achieved by sending $t\longrightarrow -\im\, \tau$
and $a_j\longrightarrow\im\, \a_j$, the metric extends smoothly onto the 
horizon at $r=r_+$ (the largest positive root of $V(r)-2M=0$) if $\tau$
is assigned the period $\beta =2\pi/\kappa$, where $\kappa$ is the surface
gravity.  Thus the Euclidean action will be given by
\bea
I_D &=& -\fft1{16\pi}\, \int_{\cal M}\sqrt{g}\,
  [R + (D-1)(D-2)\, l^{-2}]\, d^Dx - \fft1{8\pi}\,
\int_{\del{\cal M}}  \sqrt{h}\, K\, d^{D-1}x \,, \nn\\
&&= \fft{(D-1)}{8\pi\, l^2}\, 
\int_{\cal M} \sqrt{g}\, d^Dx - \fft1{8\pi}\,\int_{\del{\cal M}} 
\sqrt{h}\, K \, d^{D-1}x\,, \label{euci}
\eea
where we have included the surface contribution, whose integrand $K$ is the 
trace of the second fundamental form of the boundary, and 
the Euclidean time coordinate $\tau$ is integrated over its period 
$\beta$.

  As it stands, the expression (\ref{euci}) diverges because the volume
of the Euclidean Kerr-AdS space is infinite.  This divergence arises because
$r$ is integrated from $r_+$ to infinity.  To regularise this expression,
one terminates the $r$ integration at $r=R$, subtracts off the action for 
pure AdS adjusted so that its metric matches that of Kerr-AdS at $r=R$, 
and then sends $R$ to infinity.  This regularisation prescription is a
natural one since the Kerr-AdS metric with $m=0$ is nothing but AdS itself
(expressed in non-standard ``spheroidal'' coordinates).   Thus the regularised
action of Kerr-AdS is its action measured relative to the action of
the AdS ``vacuum'' that results from turning off the mass parameter $m$.
It should be noted that the surface terms will cancel when the background
subtraction is performed, and thus we need concentrate only on the
bulk volume integrals.

   The evaluation of the Euclidean action for the Kerr-AdS metric
$ds^2$ itself, with the radial integration running from $r_+\le r\le
R$ is straightforward.  There are two subtleties that arise in the
subtraction of the action for the AdS metric $d\bar s^2$.  Firstly, in
order to match the two metrics on the surface $r=R$ at large $R$, one
must rescale the Euclidean time coordinate $\tau$ appropriately.  Thus
the rescaled coordinate $\tau_0$ in the AdS metric is related to
$\tau$ by
\be
(V(R)- 2m)\, \tau^2 = V(R)\, \tau_0^2\,
\ee
which, for large $R$, implies
\be
\tau_0 = \Big(1- \fft{m\, l^2}{R^{D-1}}\Big)\, \tau\,,
\ee
and hence $\tau_0$ has period $\beta_0$ given by
\be
\beta_0= \Big(1- \fft{m\, l^2}{R^{D-1}}\Big)\, \beta\,.\label{beta0}
\ee

   The second subtlety concerns the integration over the volume of the
Euclideanised AdS space.  To understand this, we must look at the
relation between the coordinate system used in writing the Kerr-AdS
metric (\ref{bl}), and the coordinate system with respect to which the
Kerr-AdS metric with $m=0$ (which is just AdS) becomes the 
AdS metric written in canonical coordinates.  For general dimension
$D$ this transformation is given in \cite{gilupapo}: The 
``asymptotically-canonical'' coordinates are obtained by replacing 
$(r,\mu_i)$ by $(y,\hat \mu_i)$, where $\sum_i\hat\mu_i^2=1$ and
\be
y^2\, \Xi_i\, \hat\mu_i^2 = (r^2+ a_i^2)\, \mu_i^2\,,\label{spheroidal}
\ee
for each $1\le i\le N+\ep$.  It should be recalled that $N=[(D-1)/2]$,
$\ep= (D-1)$ mod 2, and so $D=2N+\ep +1$.  Recall also that when $D$
is even, there is no azimuthal angle $\phi_{N+1}$ associated with the
direction cosine $\mu_{N+1}$, and correspondingly $a_{N+1}\equiv 0$.
It is shown in \cite{gilupapo} that under the transformation
(\ref{spheroidal}), the Kerr-AdS metric (\ref{bl}) with $m=0$ becomes
\be
d\bar s^2 = -(1+ y^2\, l^{-2}) dt ^2 + {dy ^2 \over 1 + y^2\, l^{-2}}
  + y^2 \sum_{k=1}^{N+\ep} \bigl (
 d \hat \mu _k^2 + \hat \mu _k^2 \, d \phi _k ^2 \bigr )\,,\label{dsstat}
\ee
which can be recognised as the standard metric on AdS. 

   Expressed in terms of $(y,\hat \mu_i)$, integration over the volume
of AdS would involve integrating $y$ from 0 to some large value $y_1$.
Care is needed at the lower end of the radial
integration, $r=r_0$, where $y$ goes to zero.  From
(\ref{spheroidal}), we see that if $D=2N+2$, meaning that $\ep=1$ and
$a_{N+1}=0$, then $y=0$ just corresponds to $r=r_0=0$.  However, if
$D=2N+1$, and all $N$ of the rotation parameters $a_i$ are
non-vanishing, we shall instead have that $y=0$ corresponds to $r=r_0$
with
\be
r_0^2 = -\fft{\sum_i a_i^2\, \mu_i^2\, \Xi_i^{-1}}{\sum_j \mu_j^2\, \Xi_j^{-1}}
\,,\label{r0eq}
\ee
and this determines the lower limit of the radial integration.  (It will
be imaginary, using the radial coordinate $r$.)
If, however, the $a_i$ are not all non-vanishing, then $y=0$ will 
correspond instead to $r=0$.

   In order to perform the necessary integrations to calculate the action,
it is helpful to record the expression for the volume element of the 
Kerr-AdS metrics (\ref{bl}).  The $\mu_i$ coordinates are subject to the
constraint (\ref{muconstraint}), which can be used in order to solve
for, say, $\mu^{\phantom{\Sigma_\Sigma}}_{\sst N+\ep}$ in terms of 
the remaining $\mu_\a$, where
$1\le \a \le N+\ep-1$; \ie $\mu^{\phantom{\Sigma_\Sigma}}_{\sst N+\ep}  
=\sqrt{1-\sum_\a \mu_\a^2}$.  (Note that in even dimensions, we choose to
solve for the ``extra'' direction cosine $\mu_{N+1}$ that is not paired with
an azimuthal angle.)  
Using these coordinates, we find 
\be
\sqrt{-g} = \fft{r\, U\, 
\prod_{i=1}^N\, \mu_i}{\mu^{\phantom{\Sigma_\Sigma}}_{\sst N+\ep}\, 
                       \prod_{j=1}^N \Xi_j}\,.\label{detexp}
\ee
Thus we have
\bea
\hbox{$D= 2N+1$}: &&   \fbox{$\displaystyle \sqrt{-g} = 
      \fft{r\, U\, \prod_{i=1}^{N-1}\mu_i}{\prod_{j=1}^N \Xi_j}
   \label{sqgodd} $}\\
\hbox{$D=2N+2$}: &&  \fbox{$\displaystyle \sqrt{-g} = 
     \fft{r\, U\, \prod_{i=1}^N \mu_i}{\mu_{N+1}\, \prod_{j=1}^N\Xi_i}
      \label{sqgeven} $}\,,
\eea
where $U$ can be read off in each case from (\ref{uwline}).
It should also be noted that each direction cosine $\mu_i$ paired
with an azimuthal coordinate $\phi_i$ ranges over $0\le \mu_i\le 1$, 
but the extra unpaired direction cosine 
$\mu^{\phantom{\Sigma_\Sigma}}_{\sst N+1}$ in the case that
$D$ is even ranges over 
$-1\le \mu^{\phantom{\Sigma_\Sigma}}_{\sst N+1}\le 1$.

    The regularised action $I_D$ is now evaluated by taking the limit as
$R\rightarrow\infty$ of the difference $Y-Y_0$, where $Y$ is the action
for the Kerr-AdS metric integrated to radius $R$,
\be
Y=\fft{(D-1)\, \beta}{8\pi\, l^2}\, \int_{r_+}^R dr\, \int d^{D-2} x\, 
                    \sqrt{-g} \,,
\ee
and $Y_0$ is the action for the AdS metric integrated to radius $R$,
\be
Y_0=\fft{(D-1)\, \beta_0}{8\pi\, l^2}\, \int_{r_0}^R dr\, \int d^{D-2} x\, 
                    \sqrt{-\bar g} \,.
\ee
In these expressions, $d^{D-2} x$ represents the integration over the
direction cosines $\mu_\a$ and the azimuthal angles $\phi_i$.

   By considering specific cases, and evaluating 
\be
I_D = \lim_{R\to\infty}(Y-Y_0)\,,
\ee
we find that the Euclidean action for the $D$-dimensional Kerr-AdS metric
can be written as 
\bea
\hbox{$D$= odd}: && 
                       I_D = \fft{\beta\, {\cal A}_{D-2}}{
   8\pi\, (\prod_j \Xi_j)}\,\Big( m- l^{-2}\, \prod_{i=1}^N (r_+^2 + a_i^2) 
\Big)\,,
\label{actionoddap}
        \\
\hbox{$D$= even}: &&                       
                  I_D = \fft{{\beta\, \cal A}_{D-2}}{
   8\pi\, (\prod_j \Xi_j)}\,\Big( m -r_+\,l^{-2}\,  
\prod_{i=1}^N (r_+^2 + a_i^2)  \Big)\,,\label{actionevenap}
\eea
where $A_{D-2}$ is the volume of the unit $(D-2)$-sphere, given in
(\ref{spherevol}).
The integrations are relatively straightforward for the even-dimensional 
cases, for which the lower limit $r_0$ in the subtracted AdS action is
$r_0=0$, but it is much more complicated in the odd-dimensional cases,
due to the angle-dependent lower limit $r_0$ given by (\ref{r0eq}).  We 
have explicitly evaluated the integrals leading to (\ref{actionoddap}) for
dimensions $D=5$, 7 and 9; and the integrals leading to (\ref{actionevenap})
for $D=4$, 6, 8, 10 and 12.

   The explicit evaluation of the integrals becomes much easier in certain
special cases.  An especially simple case is when all the rotation 
parameters $a_i$ are set equal, $a_i=a$.  As discussed in \cite{gilupapo}, the
Kerr-AdS metrics in odd spacetime dimensions $D=2n+1$ can then be written
as 
\bea
ds^2 &=& -\fft{1+r^2\, l^{-2}}{\Xi}\, dt^2 + 
      \fft{U\, dr^2}{V-2m}  +
      \fft{r^2+a^2}{\Xi}\, [(d\psi+A)^2 + d\Sigma^2_{n-1}]\nn\\
&&+ \fft{2m}{U\, \Xi^2}\, [dt - a\, (d\psi+ A)]^2\,,
\eea
where 
\be
U=(r^2+a^2)^{n-1}\,,\qquad V= \fft1{r^2}\, (1+r^2\, l^{-2})(r^2+a^2)^n\,,
\ee
$\Xi\equiv 1-a^2\, l^{-2}$, and $d\Sigma_{n-1}^2$ is the standard 
Fubini-Study
metric on $\CP^{n-1}$, with K\"ahler form $J=\ft12 dA$.  The coordinate $\psi$
has period $2\pi$, and the terms $ [(d\psi+A)^2 + d\Sigma^2]$ in the
metric are nothing but the round metric on the unit sphere $S^{2n-1}$.
Following the procedure described above, and noting that in this special 
case the origin of the AdS metric occurs at $r^2=-a^2$, we find that the
action is given by
\bea
I &=& \fft{\beta\, \pi^{n-1}}{4(n-1)!\, \Xi^n\, l^2}\, \lim_{R\to\infty}\, 
[(R^2+a^2)^n -(r_+^2+a^2)^n - (1- m\, l^2\, R^{-2n})(R^2+a^2)^n]\,,\nn\\
&=& \fft{\beta\, \pi^{n-1}}{4(n-1)!\, \Xi^n}\, [m - l^{-2}\, (r_+^2 +a^2)^n]\,.
\eea
It is easily seen that this agrees with (\ref{actionoddap}), under the 
specialisation $a_i=a$.

   When all the rotation parameters are set equal in the Kerr-AdS metric in 
an even dimension $D=2n$, then, as discussed in \cite{gilupapo}, the
metric can be written as 
\bea
ds^2 &=& -\fft{\Delta_\theta\, (1+r^2\, l^{-2})}{\Xi}\, dt^2 + 
      \fft{U\, dr^2}{V-2m}  + \fft{\rho^2\, d\theta^2}{\Delta_\theta}+
      \fft{r^2+a^2}{\Xi}\,\sin^2\theta \,[(d\psi+A)^2 + d\Sigma^2_{n-2}]\nn\\
&&+ \fft{2m}{U\, \Xi^2}\, 
   [\Delta_\theta\, dt - a\, \sin^2\theta\, (d\psi+ A)]^2\,,
\eea
where
\bea
U&=&\fft{\rho^2\,(r^2+a^2)^{n-2}}{r}\,,\qquad
V= \fft1{r}\, (1+r^2\, l^{-2})(r^2+a^2)^{n-1}\,,\nn\\
\Delta_\theta &=&1- a^2\, l^{-2}\, \cos^2\theta\,,\qquad
\rho^2 = r^2 + a^2\, \cos^2\theta\,,\qquad \Xi=1-a^2\, l^{-2}\,,
\eea
and $d\Sigma_{n-2}$ is the standard Fubini-Study metric on
$\CP^{n-2}$, with K\"ahler form $J=\ft12 dA$.  Applying the procedure
described above for calculating the action, we find in this case
\bea
I&=&\fft{n!\,(2n-1)\,  2^{2n-3}\, \beta\, \pi^{n-2}}{
             (2n)!\, \Xi^{n-1}\, l^2}\, 
   \lim_{R\to\infty}[R(R^2+a^2)^{n-1} -r_+\, (r_+ + a^2)^{n-1}\nn\\
&&\qquad\qquad\qquad\qquad\qquad\qquad\qquad  -
     (1-m\, l^2\, R^{1-2n})\, R\, (R^2+a^2)^{n-1}]\,,\nn\\
&=& \fft{n!\, (2n-1)\, 2^{2n-3}\, \beta\, \pi^{n-2}}{(2n)!\, \Xi^{n-1}}\,
  [m - r_+\,\, l^{-2}\, (r_+^2+a^2)^{n-1}]\,.
\eea
This indeed agrees with (\ref{actionevenap}), specialised to the case where 
$a_i=a$.

\end{document}